\documentclass[12pt]{iopart}
\usepackage{amsfonts}
\usepackage{graphicx}
\usepackage{epsfig}
\usepackage{citesort}

\begin{document}

\title[Optical Absorption]{Independence of Optical Absorption on Auger Ionization in Single-Walled Carbon Nanotubes Revealed by Ultrafast e-h Photodoping}

\author{Mitchell D Anderson\dag, Meghan N Beattie\dag, Jack A Alexander-Webber\ddag, Robin J Nicholas\ddag, and James M Fraser\dag  
\footnote[3]{To
whom correspondence should be addressed (fraser@physics.queensu.ca)}
}

\address{\dag\ Department of Physics, Engineering Physics $\&$ Astronomy, Queen's University, Kingston, Ontario K7L 3N6, Canada}

\address{\ddag\ Department of Physics, Clarendon Laboratory, Parks Road, Oxford, OX1 3PU, UK}

\begin{abstract}
Auger-ionized carriers in a one-dimensional semiconductor are predicted to result in a strong band-gap renormalization.
Isolated single-walled carbon nanotubes (SWCNT) under high-intensity laser irradiation exhibit strong nonlinear photoluminescence (PL) due to exciton-exciton annihilation (EEA).
The presence of exciton disassociation during the rapid Auger-ionization caused by EEA would lead to a strong nonlinear absorption.
By simultaneously measuring SWCNT PL and optical absorption of isolated SWCNT clusters in the PL saturation regime, we give evidence that Auger-ionized excitons do not disassociate but remain bound.
\end{abstract}



\maketitle

\tableofcontents
\section{Introduction}

The quasi one-dimensional nature of semiconducting single-walled carbon nanotubes (SWCNTs) provides an excellent environment to study interesting photophysics arising from the strong Coulomb interaction between photo-excited carriers.
The reduced screening of optical excitations in the single rolled layer of graphene leads to an exciton binding energy which is approximately one third of the single particle bandgap \cite{Perebeinos_PRL04,Wang_Sci05}.
Consequently, excitons are stable even at room temperature where thermal energy enables exceedingly long diffusion lengths\cite{Moritsubo_PRL10,Xie_PRB12,Anderson_PRB13}.
The diffusion-reaction of excitons in SWCNTs results in many body interactions which leads to strong nonlinear optical properties\cite{Murakami_PRL09,Murakami_PRB09}, even at very low excitation fluences\cite{Xiao_PRL10}.
In this excitation regime, it was theoretically predicted that Auger-ionized carriers would give rise to a nonlinear optical response due to band gap renormalization (BGR)\cite{Konabe_NJPhys12} caused by disassociated excitons.
This prediction is supported by a recent photocurrent measurement which was attributed to the disassociation of relaxing higher energy excitons\cite{Kumamoto_PRL14}.
While SWCNTs have long been used as saturable absorbers \cite{Cho_OE09,Yim_APL08} due to their large nonlinear absorption \cite{Kamaraju_APL07,Shimamoto_APL08}, the strong nonlinear photoluminescence (PL) has been modeled only considering a reduction in quantum efficiency due to exciton-exciton interactions and not nonlinear optical absorption due to BGR.
The presence of BGR would call into question the basic understanding high fluence SWCNT PL saturation experiments\cite{Murakami_PRL09,Xiao_PRL10,Allam_PRL13} and the consequently extracted SWCNT properties\cite{Anderson_PRB13,Ishii_PRB15}.

Band gap renormalization has been previously observed in several one-dimensional systems. 
For example, BGR occurs at low density carrier doping in both highly uniform GaAs quantum wires \cite{Akiyama_SSC02} and SWCNTs \cite{Kimoto_PRB13}.
In fact, the experimentally observed BGR of carrier doped SWCNTs occurs at a much lower doping density than was theoretically predicted\cite{Spataru_PRL10}.
On the other hand, highly uniform GaAs quantum wires show no significant BGR occurs under photoexcitation (\textit{i.e.} e-h pair photodoping) \cite{Yoshita_PRB06}.
Despite this, no study has attempted to ascertain if e-h pair photodoping under high intensity photoexcitation induces the predicted BGR in undoped SWCNTs\cite{Konabe_NJPhys12}.

In this paper, we present simultaneous measurements of the optical absorption and PL of isolated clusters of several semiconducting SWCNTs in the regime of strong PL saturation.
The fluence dependence of the optical absorption and PL are measured under ultrafast ($\sim$ 150 fs) E$_{22}$ resonant optical excitation. 
We show that, despite strong PL saturation and excitation fluences far exceeding the BGR theshhold calculated by Konabe and Okada \cite{Konabe_NJPhys12}, the optical absorption remains constant, indicating the Auger process does not lead to exciton disassociation.
Additional time-resolved studies reveal that the optical absorption of a probe pulse is unaltered in the presence of a large strongly interacting exciton population, further evidencing the lack of exciton disassociation.

This paper is organized as follows.
In section 2, the methods used to measure the optical absorption and PL are briefly outlined.
Section 3 presents the results of fluences dependent PL/absorption studies and time dependent optical absorption studies.
Section 4 summarizes the present findings.

\section{Experimental methods}

Measuring the optical absorption and photoluminescence of isolated nanoparticles is a difficult feat to accomplish. 
However, using isolated individual SWCNTs or SWCNT clusters minimizes the inhomogeneous excitation effect present in large ensembles; which could obscure subtle changes to the optical absorption at low to moderate excitation intensities.
The SWCNTs used in this study were grown by HiPco and a polymer wrapping technique using poly[9,9-dioctylfluorenyl-2,7-diyl](PFO), similar to those described by Nish \textit{et al.} \cite{Nish_NatNano07}, was used to obtain a solution of highly semiconducting-enriched SWCNTs in o-xylene which was spin coated onto a quartz substrate at 2000 rpm.
The sample was chirality enriched with (8,6) SWCNTs, however some (8,7), (7,6), and (7,5) are also present.

\subsection{SWCNT Photoluminescence} 

\begin{figure}[htbp]
\centering
\includegraphics[width=0.49\textwidth]{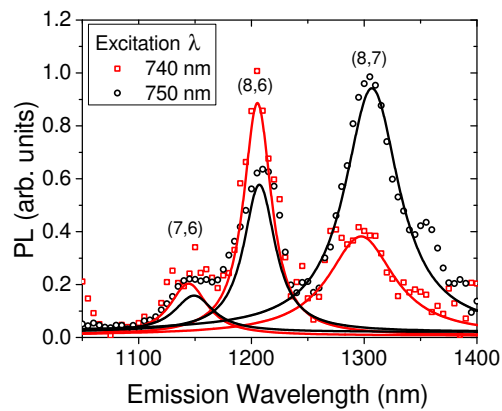}
\caption[SWCNT Bundle Photoluminescence]{PL from an isolated SWCNT cluster containing several SWCNTs excited at the E$_{22}$ transition of the (8,6) [red] and (8,7) [black] chiralities. All excitation is done using continuous-wave light.}
\label{fig:PLWVS}
\end{figure}

In order to locate and study the isolated SWCNT clusters (containing $\sim$ 5 SWCNTs), a Ti:Sapphire laser was used running in continuous wave (CW) mode and tuned resonant to the E$_{22}$ excitonic transition.
Optically excited E$_{22}$ excitons quickly relax to E$_{11}$ excitons by coupling to phonon modes\cite{Manzoni_PRL05}.
The isolated SWCNT clusters were located by raster scanning the sample while it was being illuminated with a $\sim$ 2 $\mu$m excitation spot, created by focusing the excitation light through an aspheric lens (NA $\sim$ 0.55).
The PL is collected by the same lens that focuses the excitation light and is separated from the main beam path using a dichroic mirror.
A LN$_{2}$ cooled InGaAs single element detector and a spectrometer were used to measure PL at the E$_{11}$ excitonic transition.
The wavelengths were chosen at the excitation and emission frequencies of the (8,7) SWCNT due to experimental constraints.
Selection of the SWCNT orientation was performed with a set of two linear polarizers: one polarizer to control the polarization of the excitation light and a second polarizer to analyze the polarization of the PL.
Generally, in addition to the (8,7) emission peak, the localized bright SWCNT clusters had emission peaks corresponding to (7,6) and (8,6) chiralities, as seen in Figure: \ref{fig:PLWVS}.
The E$_{11}$ emission linewidths around 25-40 meV, consistent with inhomogeneous broadening of the encapsulated samples.
It can be seen from Figure: \ref{fig:PLWVS} that, even when the (8,7) SWCNT is targeted for excitation, there is still significant PL from the (8,6) SWCNT.
However, the strong increase in the (8,7) SWCNT PL peak on resonance indicates that there is not significant energy transfer between the SWCNT chiralities.
One of the consequences of the SWCNT clustering is that careful attention has to be paid to the entire integrated PL when comparing PL saturation.
This is to ensure that the power dependent absorption signal is not being significantly affected by the non-targeted chiralities.

Since a SWCNT emits light preferentially polarized along its axis, the orientations of the SWCNTs are found by measuring polarization of the PL using a polarization analyzer.
The polarization analyzer works by rotating a linear polarizer in the PL path.
Figure: \ref{fig:WVSpolfig} (b) shows an example of the PL polarization from a SWCNT cluster. 
Despite the presence of several SWCNTs, the PL is moderately polarized indicating that the SWCNTs are reasonably well aligned. 
This is exploited in the technique we use to measure the optical absorption.
Since the SWCNTs are located using an excitation polarization in the horizontal direction the resulting SWCNT PL is close to the horizontal direction.
During experimentation, PL from all SWCNTs peaks are measured to ensure the saturation of a single (on resonant) SWCNT chirality is not negligible compared to the total PL signal (\textit{i.e.}, PL from all SWCNTs saturate).

\subsection{SWCNT Optical Absorption}

\begin{figure}[h!]
\centering
\includegraphics[width=0.6\textwidth]{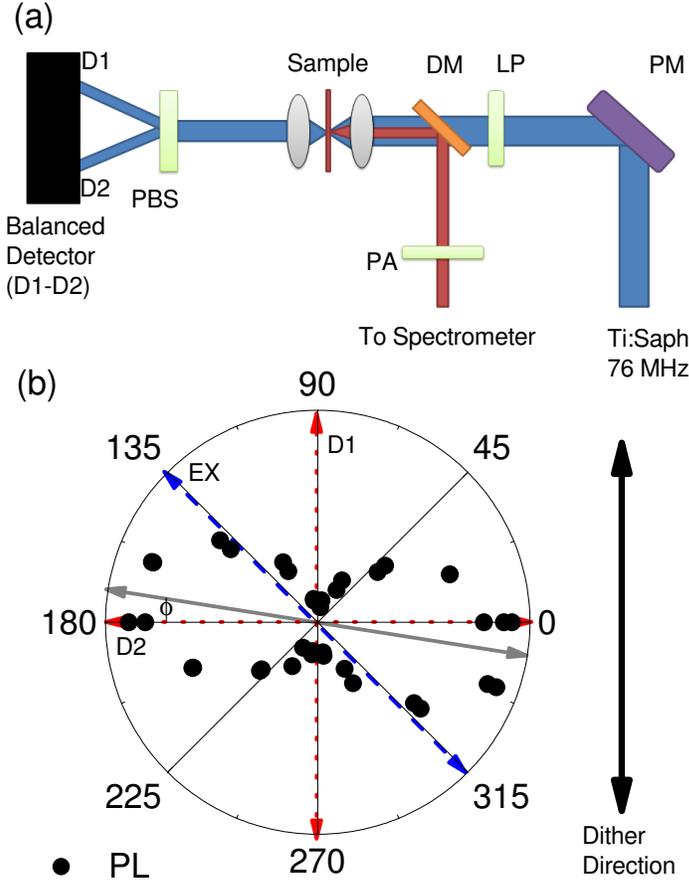}
\caption[SWCNT Polarization Diagram]{(a) Experimental design including: a piezo modulated mirror (PM), a linear polarizer (LP), a dichroic mirror (DM), and a polarization analyzer (PA). (b) Orientation of the polarizations sent to detector 1 (D1) and detector 2 (D2). Also, the excitation polarization (EX) and a sample data set (black dots) at an angle $\phi$ from the horizontal axis. The dither direction is across the SWCNT axis (black arrows). SWCNT clusters are chosen which emit close to the horizontal axis. Despite being clustered, the SWCNTs exhibit fairly polarized PL indicating they are aligned and straight.}
\label{fig:WVSpolfig}
\end{figure}

While it is reasonably straightforward to measure the PL from isolated SWCNTs \cite{Lefebvre_PRL03}, it is much more difficult to directly measure the optical absorption.
Only recently, have direct measurements of the optical absorption of individual SWCNTs been achieved using two different techniques: polarized light microscopy\cite{Lefebvre_NR11} and spatial modulation\cite{Blancon_NatComm13,Oudjedi_JPCL13,Christofilos_JPCL12}.
In this work, both of these techniques are adapted and combined to achieve SWCNT absorption measurements (illustrated in Figure: \ref{fig:WVSpolfig}).
These techniques rely on the control or characterization of the SWCNT orientation.
Since the spin coated samples have SWCNTs in all orientations, the SWCNTs are post selected using the orientation as characterized by PL measurements (see Figures: \ref{fig:PLWVS} (b) and \ref{fig:WVSpolfig}).
Once the SWCNT orientation is known, polarization microscopy can be used to differentiate SWCNT absorption from particulate absorption.

\subsubsection{Polarization Microscopy}

The optical absorption/scattering of the E$_{11}$, E$_{22}$, \textit{etc} absorption peaks of a SWCNT are strongly polarization sensitive \cite{Islam_PRL04}.
Only light which is polarized along the SWCNT axis is absorbed.
Since a SWCNT is very weakly absorbing and the SWCNT sample may contain any number of particulates which absorb light much more strongly, a measurement technique that is sensitive to polarization specific optical absorption is very useful.
Therefore, in order to strongly bias the optical absorption signal to only measure SWCNT absorption, polarization microscopy is used.

Polarization sensitive measurements are performed using: a linear polarizer, to adjust the excitation polarization 45$^{\circ}$ from the SWCNT axis, and a Wollaston prism, to separate the vertical and horizontal components of the light.
The two polarizations are then sent to two arms of a balanced amplified silicon photodetector (BAP).
The BAP amplifies and outputs the differential signal of the two detectors (\textit{i.e.} the vertical and horizontal polarization components of the excitation beam).
Using a BAP has the added benefit of canceling the majority of noise caused by laser fluctuations.
The absorption experiment is designed such that the maximum signal is obtained for SWCNTs oriented along one of the principle axes (\textit{i.e.,} the x (horizontal) or y (vertical) axis).
For experimentation, the SWCNTs are preferentially selected in the horizontal direction by controlling the excitation polarization.
Only relative absorption is important for this study, so the depolarization effect of the high NA focusing lens can be ignored\cite{Herman_APL12}. 
The current implementation of polarization microscopy is able to reduce the relative noise to signal ratio to $10^{-4}$. 
However, to observe the weak SWCNT absorption signal a background reduction of $10^{-6}$ is required.
For this reason, an additional technique to reduce background noise is necessary.

\subsubsection{Dithered Signal Demodulation}

To measure the exceedingly small absorption signals, the signal is modulated by dithering the excitation beam across the SWCNT axis.
This is done by using a piezo mirror dithering at 1KHz.
To calculate the demodulated signal, the time dependent transmitted power, $\rm{P}_{T}(t)$, is used.
This is given by $\rm{P}_{T} = \rm{P}_{I} - \sigma \int  \rm{I}(x,y(t)) \rm{dx}$, where $\rm{P}_{I}$ is the incident power. 
The time-dependent intensity is modeled using a Gaussian spatial profile, $ \rm{I}(x,y(t)) = \rm{I}_{0}\exp[(-2(x^{2}+y(t)^{2})/\omega_{0}^{2}]$ where $y(t)=y_{0}+\delta \sin(2\pi f t)$.
The spatial modulation depends on the driving frequency (f) and the dither distance ($\delta$). 
The peak intensity is given by $\rm{I}_{0} = 2 \rm{P}_{I}/\pi\omega_{0}^{2}$.
The absorption coefficient is defined as $\sigma = \sigma_{abs} \times N_{C}$, where $\sigma_{abs}$ is the absorption cross-section of the carbon atom and $N_{C}$ is the number of carbon atoms per unit length.
The number of carbon atoms per unit length calculated using $N_{C}=N_{C}^{cell}/L_{cell}$. 
 
Using a lock-in amplifier to detect the signal, the demodulated power can be calculated using,

\begin{equation}
	\rm{P}_{demod} = -\frac{\sigma}{T}\int \rm{dx} \int\limits_{0}^{T} \rm{dt}   \rm{I}(x,y(t)) \sin(2\pi f' t + \phi).
\end{equation}
Assuming $\delta$ is small, the Taylor expansion of the integrand about $\delta = 0$ can be used. 
The maximum 1f and 2f demodulated signals (assuming the excitation spot is centered horizontally along the SWCNT and scanned vertically) are calculated to be,

\begin{eqnarray}
	\frac{\rm{P}_{1f}(y)}{\rm{P}_{I}} &=& \sigma  C_{S} \sqrt{8\delta^{2}}  y \exp \left( -2\frac{y^{2}}{\omega_{0}^{2}} \right) \label{eq:1fdemod}, \\
		\frac{\rm{P}_{2f}(y)}{\rm{P}_{I}} &=& \sigma C_{S}\delta^{2}  \left[ 1- 4\left(\frac{y}{\omega_{0}}\right)^2 \right] \exp\left({-2\frac{y^{2}}{\omega_{0}^2}}\right), \label{eq:2fdemod}
	\label{eq:demod}
\end{eqnarray}
where $C_{S}=\rm{erf}(L/\sqrt{2}\omega_{0})/(\sqrt{\pi}{\omega_{0}^3})$ is a constant which depends on the spatial extent of the SWCNT (L) and the spot size ($\omega_{0}$).

The 1f and 2f demodulated signals both have advantages and disadvantages.
The advantage of the 1f demodulated signal is its larger signal caused by the scaling with $\delta$ and not $\delta^{2}$.
The 2f demodulated signal has the advantage that it is non-zero when the excitation spot is centered on the SWCNT.
This allows for studies to be performed with the excitation spot fixed centrally along the SWCNT axis.

\subsubsection{Measuring Optical Absorption}

\begin{figure}[htbp]
\centering
\includegraphics[width=0.45\textwidth]{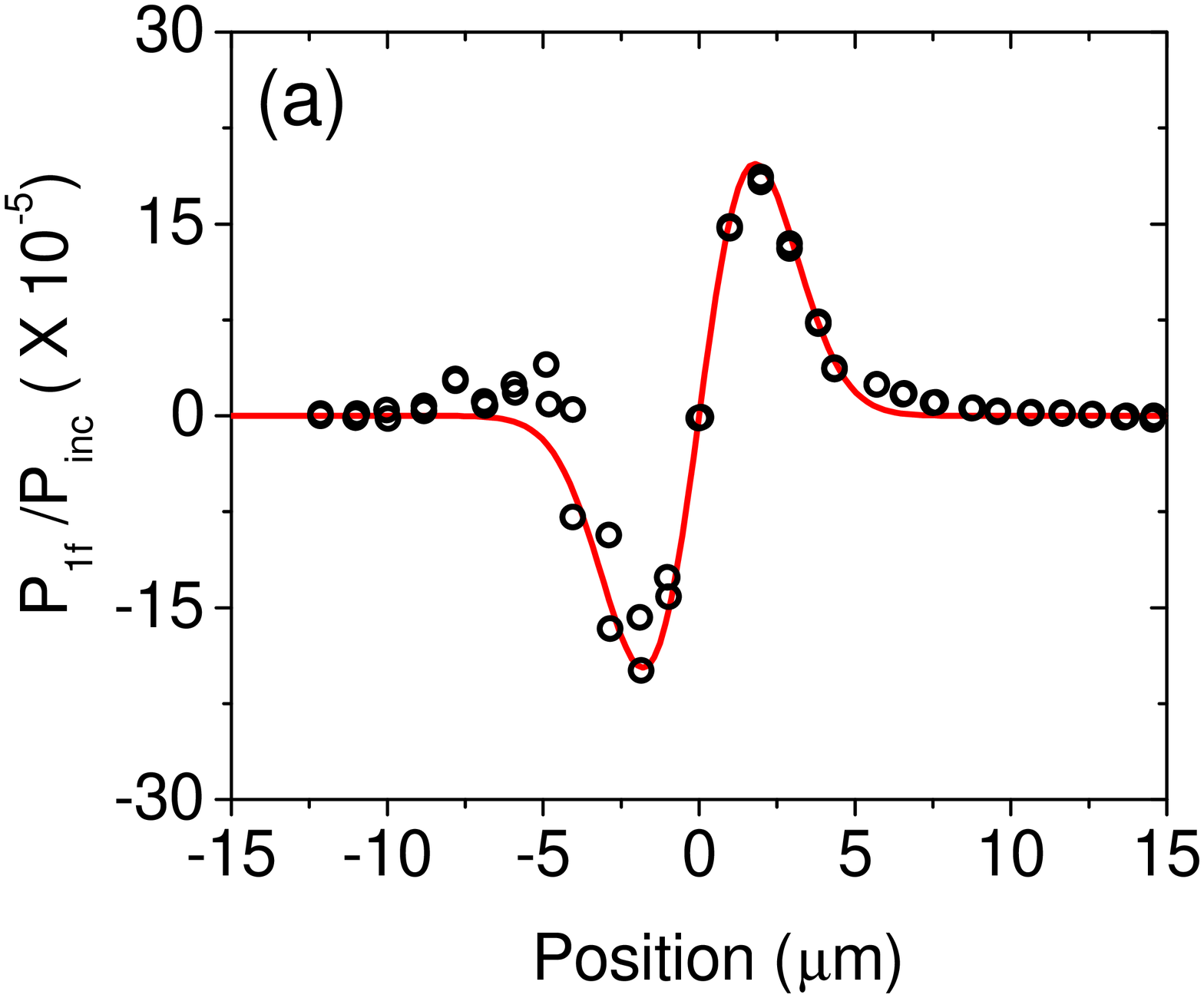}
\includegraphics[width=0.45\textwidth]{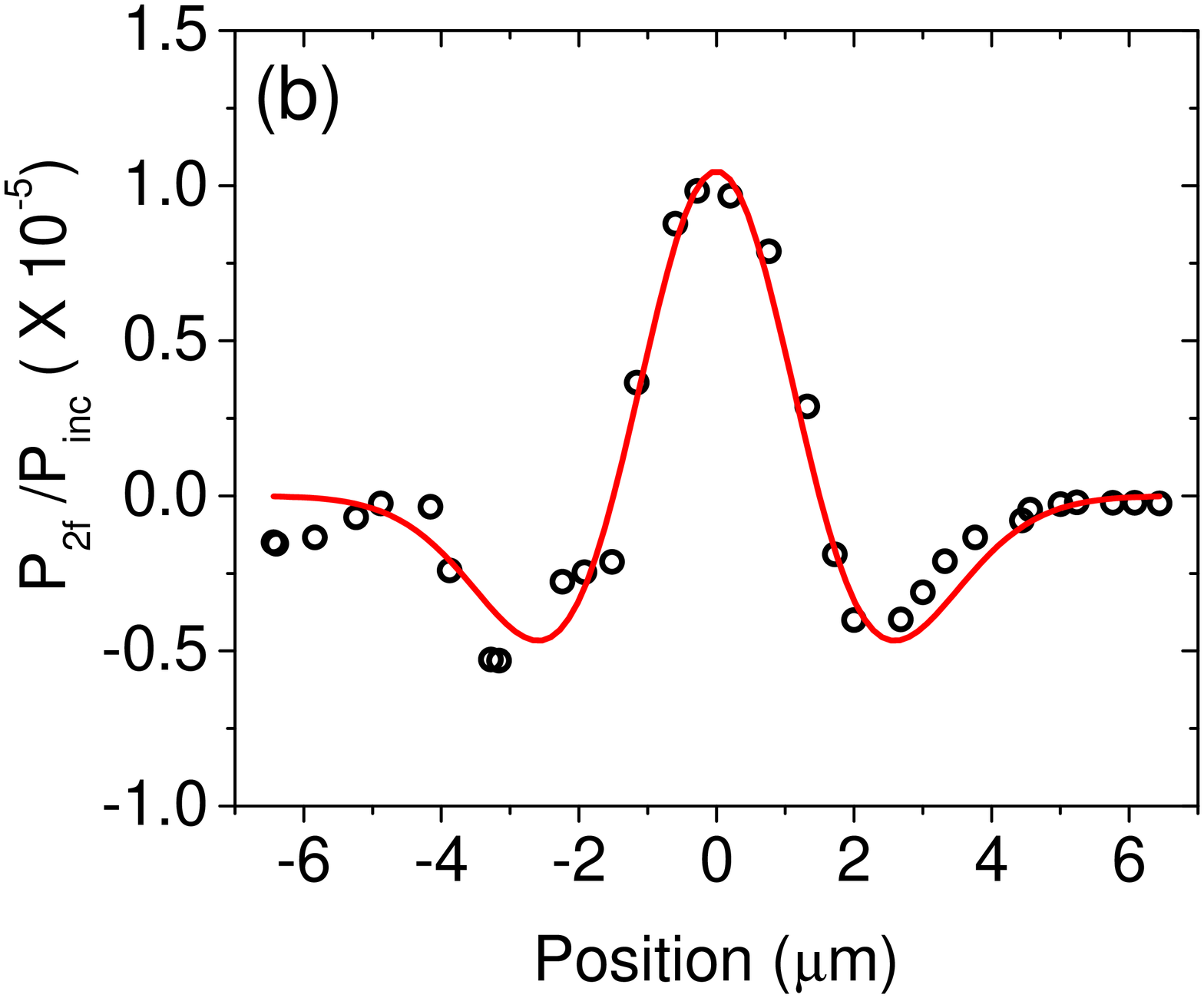}
\caption[Absorption Profiles]{A 1D profile of a SWCNT bundle demodulated at (a) 1f (fit with Equation: \ref{eq:1fdemod}) and (b) 2f (fit with Equation: \ref{eq:2fdemod}). The absorption of the SWCNT bundle is proportional to $\rm{P}_{demod}/\rm{P}_{I}$. As expected, the measured 1f signal is larger than the 2f signal.}
\label{fig:absprof}
\end{figure}

Two methods are used to measure the optical absorption of a SWCNT cluster.
Fluence-dependent SWCNT measurements are performed by taking 2D spatial scans of the SWCNT clusters.
In these studies, 1f demodulation is used which has a high signal to noise ratio and can be fit using Equation: \ref{eq:1fdemod}.
This eliminates the fluence-dependent background which manifests as a DC offset.
Since many of the parameters needed to extract real numbers (\textit{e.g.} length, number of nanotubes, etc.) are unknown, the peak absorption signal is set to one. 
The important extracted factor is the relative absorption ($\sigma \propto \rm{P_{demod}}/\rm{P}_{I}$), which is proportional to the peak of the absorbed power over the incident power.
An example of a 1D slice of the SWCNT cluster, raster scanned perpendicular to the SWCNT axis and demodulated at 1f, is shown in Figure: \ref{fig:absprof} (a).

Time-resolved PL measurements are performed using two beam paths: one that is fixed and a second which is delayed in time.
Femtosecond excitation correlation spectroscopy \cite{Hirori_PRL06,MiyauchiFEC_PRB09,Xiao_PRL10,Anderson_PRB13} is used to measure the relaxation of the exciton population while the cumulative optical absorption is measured.
Since the total power of the optical excitation is constant, the background remains constant and 1D scans are unnecessary. 
To record the optical absorption, a single 1D scan of the SWCNT is performed demodulating at 2f (\textit{e.g.} Figure: \ref{fig:absprof} (b)) and is fit with Equation: \ref{eq:2fdemod}.
The excitation spots are then centered at the peak of the absorption signal, which also corresponds to the peak of the PL signal, and the PL and absorption signals are measured simultaneously as a function of the pulse delay.



\section{Results and discussion}

\subsection{SWCNT Bundle Photoluminescence Saturation} 

The saturation of SWCNT PL has been observed in various SWCNT samples including: air-suspended \cite{Xiao_PRL10}, encapsulated ensemble \cite{Murakami_PRL09}, and encapsulated single \cite{Santos_PRL11}.
Here, isolated encapsulated SWCNT clusters are studied which would be similar to those studied by Santos \textit{et. al.} \cite{Santos_PRL11}.
Since Santos \textit{et. al.} observed rapid defect induced PL quenching and unreproducible PL saturation curves, special care is taken to select data which was reproducible.
Any data taken after the SWCNT was noticeably damaged was discarded.
For some SWCNTs, irreversible changes in the PL were observed beyond $\sim$ 90 $\mu$W, indicating that the SWCNT(s) or their encapsulation were damaged.
Other SWCNTs were studied up to $\sim$ 1 mW without damage.
Using an absorption cross-section of 522 nm$^{2}$/$\mu$m for an (8,7) HiPCO SWCNT \cite{Vialla_PRL13}, this would correspond to an upper electron-hole density of 0.3 nm$^{-1}$, approaching the Mott density.

\begin{figure}[htbp]
\centering
\includegraphics[width=0.6\textwidth]{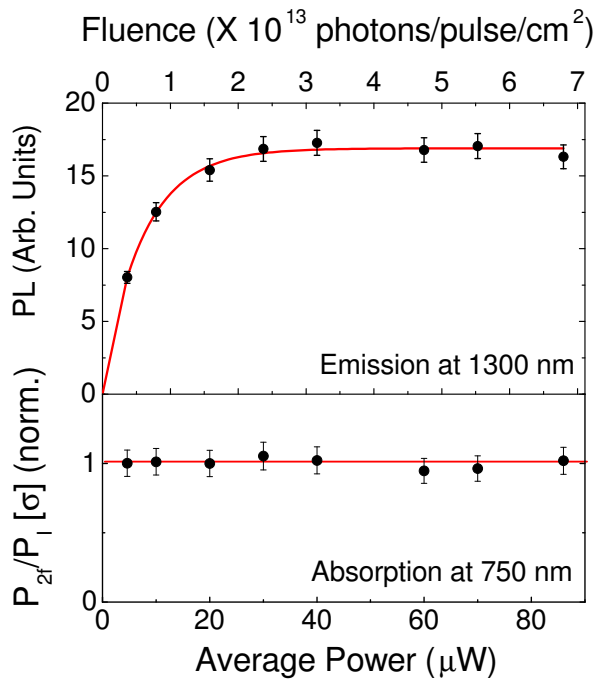}
\caption[SWCNT bundle saturation with absorption]{PL saturation of (8,7) SWCNT(s) with corresponding relative optical absorption ($\sigma$). The strong non-linearity in PL is not accompanied by a decrease in the optical absorption. Red trend lines are included showing PL saturation and linear absorption. This indicates that the PL saturation is dominated by processes occurring to the E$_{11}$ excitons (such as diffusion-reaction). All error bars are standard deviation.}
\label{fig:satandabs}
\end{figure}

In these SWCNTs disorder diffusion, not intrinsic diffusion, is expected resulting in significantly shorter diffusion lengths than those observed in air-suspended SWCNTs \cite{Anderson_PRB13}.
A consequence of the shorter diffusion length is evidenced in Figure: \ref{fig:satandabs}, which has a PL saturation point $\sim$ 10$^{13}$ photons/pulse/cm$^{2}$.
This is in contrast to the observed PL saturation point of $\sim$ 10$^{11}$ photons/pulse/cm$^{2}$ in air-suspended SWCNTs\cite{Xiao_PRL10}, which Konabe and Okada \cite{Konabe_NJPhys12} suggest is sufficient to cause BGR and nonlinear optical absorption.
Despite the two orders of magnitude higher fluence, the onset of PL saturation is not accompanied by a decrease in the optical absorption.
This gives strong evidence that the observed PL saturation does not occur due to a non-linearity in the optical absorption as suggested by Konabe and Okada \cite{Konabe_NJPhys12}.

\subsection{Time-Resolved Optical Absorption Studies} 

\begin{figure}[htbp]
\centering
\includegraphics[width=0.45\textwidth]{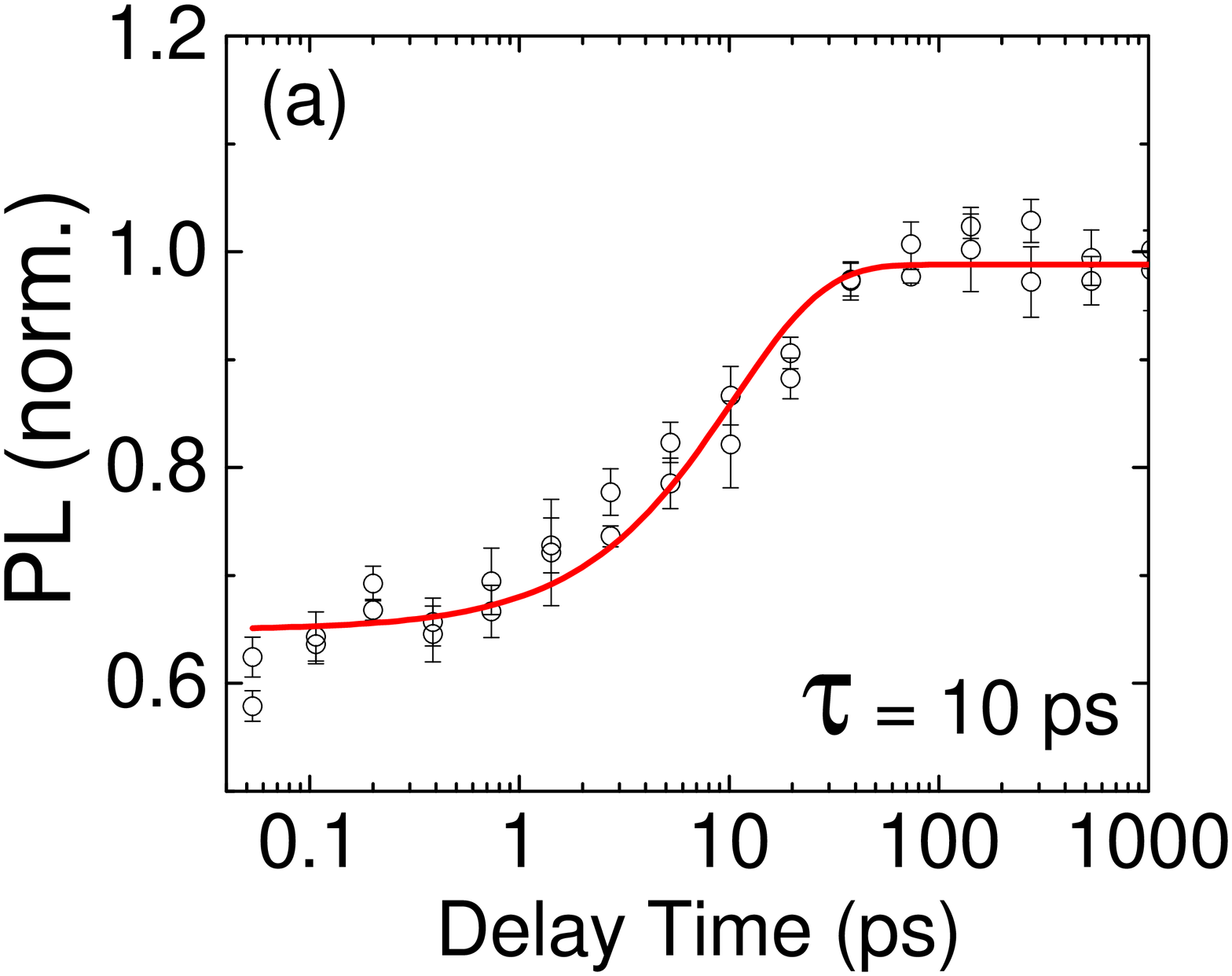}
\includegraphics[width=0.45\textwidth]{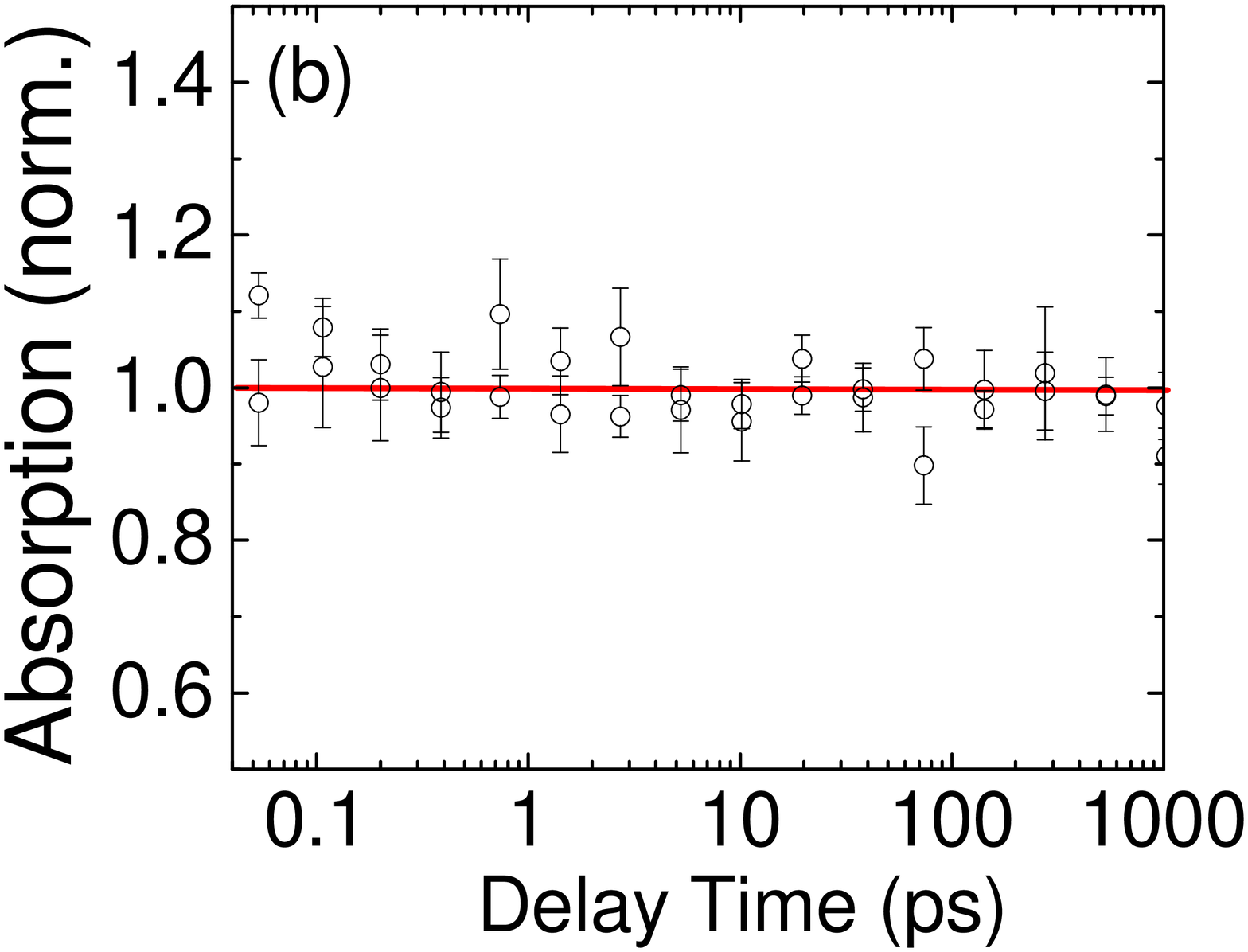}
\caption[Time-resolved PL and Absorption]{(a) PL relaxation of an isolated SWCNT bundle. The effective lifetime is 10 ps (FEC signal fit using a mono-exponential, see Ref: \cite{Anderson_PRB13}), significantly shorter than the air-suspended SWCNTs.  (b) The corresponding optical absorption, which corresponds to the sum of the optical absorption of the probe pulse and delayed pump pulse. Absorption data collected using a 2f demodulated signal. All errors are standard deviations.}
\label{fig:FECABSPL}
\end{figure}

Time-resolved PL and absorption studies were performed to test if the optical absorption was affected by the excited state of the SWCNT.
To do this, FEC scans were performed which measured the relaxation time of the SWCNT bundle.
In this study, both the optical absorption and the PL measured as the sum signal of the two excitation pulses.
Figure: \ref{fig:FECABSPL} (a) gives a typical FEC signal demonstrating the SWCNT exciton population relaxation \cite{Xiao_PRL10,Anderson_PRB13,Xiao_PRB14}.
The relaxation lifetime ($\tau$) of this SWCNT cluster was fit using the FEC signal assuming a mono-exponential decay, $FEC\propto \int_{0}^{\tau'}\exp (-t/\tau)dt$, where $\tau'$ is the delay time (see Ref: \cite{Anderson_PRB13}). The measured relaxation lifetime was 10 ps, significantly shorter than typical exciton population relaxation times observed in longer pristine SWCNT samples \cite{Berciaud_PRL08,Anderson_PRB13,Xiao_PRB14}, as expected due to environmental perturbations and end quenching.

The time-dependent optical absorption is measured using the 2f demodulated signal.
The excitation spots are centered on the SWCNT.
Since both the probe pulse and the delayed pump pulse are modulated, the total absorption signal is detected.
If the presence of excitons reduces the optical absorption of the second pulse, a reduction of the total absorption signal is expected (similar to that observed in the PL signal).
Two scans are taken, one increasing the delay time and the second decreasing the delay time. 

Figure \ref{fig:FECABSPL} shows that, in contrast to the observed PL relaxation, there is no observable change in the SWCNT optical absorption over the range of 200 fs to 1 ns.
This provides evidence that a strongly populated E$_{11}$ exciton state does not significantly affect the optical absorption of E$_{22}$ excitons.
This indicates that, like in GaAs quantum wires, the presence of e-h pair photodoping does not cause BGR.

\section{Summary}

Isolated SWCNT clusters (with $\sim$ 5 SWCNTs) were studied using simultaneous measurements of the optical absorption and photoluminescence. 
Results provide evidence counter to the theoretical prediction that under high optical excitation, carrier doping due to exciton disassociation would lead to BGR reducing the optical absorption.
The PL saturation for the encapsulated SWCNT bundles occurred at fluences two orders of magnitude higher than air-suspended SWCNT samples.
In this high fluence regime, the observed hard saturation of PL from the SWCNT cluster was not accompanied by a decrease in the observed absorption cross-section.
In addition, time-resolved PL shows that there was no observable change in the absorption cross-section as the exciton population decayed.
This indicates that for e-h pair photodoping, at levels sufficient to fully saturate the PL, Auger recombination of excitons does not result in exciton disassociation, which would lead to a BGR.

\ack

The authors would like to thank M. Abdelqader and R. Martel for helpful discussions.
This work is funded by the Natural Sciences and Engineering Research
Council of Canada, the Canadian Foundation for Innovation, and the
Ministry of Economic Development and Innovation (Ontario).

\section*{References}
\providecommand{\newblock}{}


\end{document}